\documentclass[useAMS,usenatbib]{mn2e}
\usepackage{epsfig}
\usepackage{threeparttable}
\usepackage{amsmath}

\title[On the progenitors of MSPs by the recycling evolutionary channel]{On the progenitors of millisecond pulsars by the recycling evolutionary channel}
\author[W. M. Liu, and W. C. Chen]{Wei-Min Liu$^{1}$, and Wen-Cong Chen$^{1,2,3}$\thanks{E-mail:
chenwc@nju.edu.cn},\\ $^1$ Department of Physics, Shangqiu Normal University, Shangqiu 476000, China;\\
 $^2$ School of Physics and State Key Laboratory of Nuclear Physics and Technology, Peking University, Beijing 100871, China;\\
 $^3$ Key Laboratory of Modern Astronomy and Astrophysics (Nanjing University), Ministry of Education, Nanjing 210093, China }
\begin{document}

\date{}

\pagerange{\pageref{firstpage}--\pageref{lastpage}} \pubyear{2011}

\maketitle

\label{firstpage}

\begin{abstract}
The recycling model suggested that low-mass X-ray binaries (LMXBs)
could evolve into binary millisecond pulsars (BMSPs). In this
work, we attempt to investigate the progenitor properties of BMSPs
formed by the recycling evolutionary channel, and if
sub-millisecond pulsars can be produced by this channel. Using
Eggleton's stellar evolution code, considering that the dead
pulsars can be spun up to a short spin period by the accreting
material and angular momentum from the donor star, we have
calculated the evolution of close binaries consisting of a neutron
star and a low-mass main-sequence donor star, and the spin
evolution of NSs. In calculation, some physical process such as
the thermal and viscous instability of a accretion disk, propeller
effect, and magnetic braking are included. Our calculated results
indicate that, all LMXBs with a low-mass donor star of 1.0 - 2.0
$M_\odot$ and a short orbital period ($\la 3-4 \rm d$) can form
millisecond pulsars with a spin period less than 10 ms. However,
it is difficult to produce sub-millisecond pulsars by this
evolutionary channel. In addition, our evolutionary scenario
cannot account for the existence of BMSPs with a long orbital
period ($P_{\rm orb}\ga 70-80\rm d$).
\end{abstract}


\section{Introduction}

Millisecond pulsars (MSPs) and normal pulsars have distinct
observed properties, and they inhabit two different regions in
magnetic field - spin period ($B - P$) diagram \citep{man05}.
Normal pulsars have a spin period of $P\sim 1 ~\rm s$ and a
magnetic field of $B\sim 10^{12}~\rm G$. However, MSPs show some
distinct observed properties such as short spin period ($P \la \rm
20~ ms$), low spin-down rate ($\dot{P}\sim 10^{-19}-10^{-21}\rm~
s\,s^{-1}$), old characteristic age
($\tau=P/(2\dot{P})\sim10^{9}-10^{10}\rm ~yr$), and weak surface
magnetic fields ($B\sim10^{8}-10^{9}\rm~ G$) \citep{man04,lori08}.
About 75\% MSPs are in binary system (called binary millisecond
pulsars, BMSPs), whereas that is only $\la 1\%$ for  normal
pulsars.

At present, there exist two scenarios to account for the formation
of MSPs. The first one is the recycling model, in which MSPs are
proposed to be the evolutionary product of neutron star (NS)
low-mass X-ray binaries (LMXBs) or intermediate-mass X-ray
binaries (IMXBs) \citep{alpa82}. The pulsar crossed the so-called
deathline accretes the mass and angular momentum from the donor
star that overflows its Roche lobe, and can be subsequently spun
up to a millisecond spin-period \citep{bhat91,taur06}. During
accretion, the magnetic field of the NS decrease to be $B\sim
10^{8}-10^{9}~\rm G$ due to accretion-induced field decay
\citep{kona97}. When the mass transfer ceases, a BMSP consisting
of a recycling NS and a low-mass ($\la 0.4 M_{\odot}$) helium
white dwarf is produced. The discovery of the accreting
millisecond X-ray pulsar Sax J 1808.4-3658 presented strong
support to this scenario \citep{wijn98}. Recent optical
observations also confirm that there exists a transition link
between X-ray pulsar and millisecond radio pulsar
\citep[e.g.][]{arch09}.

In another evolutionary channel, MSPs may be formed by
accretion-induced collapse (AIC) of ONeMg white dwarfs
\citep{mich87}. When the mass of an ONeMg white dwarf reaches the
Chandrasekhar mass limit by accreting from its donor star, the
electron-capture process leads to a gravitational collapse rather
than a Type Ia explosion, and results in the formation an NS
\citep{nomo91} \footnote{When the ONeMg core of an asymptotic
giant branch star \citep{sies07,poel08} or a He star
\citep{nomo87} grow to a critical mass, electron-capture supernova
can also produce an NS. \cite{pods04} suggested that the minimum
mass of the NS progenitor may be $10-12~M_{\odot}$ for single
stars, while this value can be $6-8~M_{\odot}$ in binaries.}. If
MSPs formed by the collapse of low field ($10^{3}-10^{4}$ G) white
dwarf population \citep{jord07}, their magnetic field should be in
the range $10^{8}-10^{10}$ G, and without invoking significant
field decay. Recently, the calculated birthrates by population
synthesis approach indicate that the AIC channel may play an
important role in forming MSPs (Hurley et al. 2010). As an
alternative formation of MSPs, this evolutionary channel has been
widely explored by some authors \citep{wic09,hurl10,chen11c}. In
particular, \cite{du09} argued that AIC process of massive white
dwarfs can produce sub-millisecond pulsars (quark stars) with a
spin-period less than 1 ms (or less than 0.5 ms).

The purpose of this paper is to systemically explore  the initial
parameter space of LMXBs that could evolve into BMSPs via the
recycling evolutionary channel. In addition, we also attempt to
examine if this channel can form the so-called sub-millisecond
pulsar. The structure of this paper is as follows. We describe the
input physics that is necessary in the evolution calculation of
LMXBs in section 2. The calculated results are presented in
section 3. Finally, we give a brief discussion and summary in
section 4.

\section{Input physics}

Using a stellar evolution code developed by Eggleton (see Eggleton
1971, 1972, 1973), which has been updated with the latest input
physics over the past three decades \citep{han94,pol95,pol98}, we
calculate the evolution of binaries consisting of a NS (of mass $
M_{\rm NS}$) and a main-sequence donor star (of mass $M_{\rm d}$)
 \footnote{Certainly, NSs may also recycled by accreting the material from
the He star companion. However, some studies show that the
evolution products of NS + He star systems should be
intermediate-mass binary pulsars or high-mass binary pulsars
\citep[see][]{fra02,chen11b}. }, and test if they can evolve into
MSPs. The stellar OPAL opacities was taken from \cite{roge92} and
\cite{alex94} for a low temperature. In our calculation, the ratio
of mixing length to local pressure scale height was set to be
$2.0$, and the overshooting parameter of the donor star (with a
solar chemical composition $X = 0.70$, $Y = 0.28$, and $Z = 0.02$)
is taken to be 0 \citep{dewi02}.

\subsection{Accretion disk instability}
With nuclear evolution, the donor star overflows its Roche lobe,
and transfer hydrogen-rich material onto the NS. Due to the high
angular momentum, the accreting material forms a disk surrounding
the NS. If the effective temperature in the accretion disk is
below $\sim6500~\rm K$ (the hydrogen ionization temperature), the
disk accretion should be thermally and viscous unstable
\citep{para96,kin97,las01}. Meanwhile, the accreting NS will be a
transient X-ray source, which appears as short-lived outbursts
phase and long-term quiescence phase. Recently, \citet{chen11a}
found that accretion disk instability model successfully
reproduces the orbital period and the mass of the WD of PSR
J1713+0747.

When the mass transfer rate $-\dot{M}_{\rm d}$ is lower than the
critical mass-transfer rate \citep{para96,dubu99}
\begin{equation}
\begin{aligned}
&\dot{M}_{\rm cr}\simeq 3.2\times10^{-9}{\left(\frac{M_{\rm
NS}}{1.4~{M_\odot}}\right)}^{0.5}{\left(\frac{M_{\rm d}}{1.0~{M_\odot}}\right)}^{-0.2}\\
&~~~~~~~~~{\left(\frac{P_{\rm orb}}{1.0~\rm
d}\right)}^{1.4}~~{M_\odot~\rm yr^{-1}},
\end{aligned}
\end{equation}
where $P_{\rm orb}$ is the orbital period of the binary, the NS
accretes only during outbursts. Defining a duty cycle $d$ to be
the ratio of the outburst timescale to the recurrence time
\footnote{\citet{kin03} proposed that the typical value of duty
cycle is about 0.1 to 0.001. In this work, we take $d=0.01$.}, the
accretion rate of the NS $\dot{M}_{\rm ac}=-\dot{M}_{\rm d}/d$.
Otherwise for a high mass transfer rate $-\dot{M}_{\rm
d}>\dot{M}_{\rm cr}$, we assume $\dot{M}_{\rm ac}=-\dot{M}_{\rm
d}$. Certainly, the mass growth rate of the NS should suffer the
limitation of the Eddington accretion rate ($\dot{M}_{\rm
Edd}\approx1.5\times10^{-8}M_\odot$). The excess material is
assumed to be expelled from the vicinity of the NS by radiation
pressure, and carries away the specific orbital angular momentum
of the NS.

\subsection{Magnetic braking}
Low-mass donor star would be braked to spin down by the coupling
between the magnetic field and the stellar winds \citep{verb81}.
However, the tidal interaction between the donor star and the NS
would continuously spin the star back up co-rotation with the
orbital rotation \citep{patt84}. Therefore, magnetic braking
mechanism indirectly carries away the orbital angular momentum of
binaries.

For the angular momentum loss rate via magnetic braking,
\citet{rapp83} developed an empirical formula, i. e.
\begin{equation}
\dot{J}_{\rm mb}\simeq -3.8\times
10^{-30}M_2R_{\odot}^4(R_2/R_{\odot})^{\gamma}\omega^3\,{\rm
dyn\,cm},
\end{equation}
where $R_{2}$ is the radius,  $\omega$ the angular velocity of the
donor star, and $\gamma$ is a dimensionless parameter in the range
of zero to four. This standard magnetic braking model is widely
applied in studying the evolution of cataclysmic variables.
However, studies on rapidly rotating low-mass stars with a spin
period below 2.5 - 5 days in young open clusters show that the
standard model overestimates the angular momentum loss rate
\citep{quel98,andr03}.

In calculation, we adopt an induced magnetic braking description
given by \citet{sill00}, in which the angular momentum loss rate
is
\begin{equation} \dot{J}_{\rm mb}=
\left\{\begin{array}{l@{\quad}l} -K\omega ^{3}\left(\frac{R_{\rm
d}}{R_{\odot}}\frac{M_{\odot}}{M_{\rm d}}\right)^{1/2},
& \omega\leq\omega_{\rm crit} \strut\\
-K\omega \omega_{\rm crit}^{2}\left(\frac{R_{\rm
d}}{R_{\odot}}\frac{M_{\odot}}{M_{\rm d}}\right)^{1/2} , &
\omega>\omega_{\rm crit} \strut\\\end{array}\right.
\end{equation}
where $K=2.7\times 10^{47} \rm g\, cm^{2}$ \citep{andr03},
$\omega_{\rm crit}$ is the critical angular velocity at which the
angular momentum loss rate reaches a saturated state,
$\omega=2\pi/P_{\rm orb}$ and $R_{\rm d}$ are the angular velocity
and the radius of the donor star, respectively. \citet{kim96}
proposed that $\omega_{\rm crit}$ is inversely proportional to the
convective turbulent timescale of the star when its age is 200 Myr
, i. e.
\begin{equation}
\omega_{\rm crit}=\omega_{\rm
crit,\odot}\frac{\tau_{\odot}}{\tau},
\end{equation}
where $\omega_{\rm crit,\odot}=2.9\times 10^{-5}$ Hz,
$\tau_{\odot}$, and $\tau$ are the convective turbulent timescales
of the Sun and the donor star, respectively.

\subsection{Spin evolution of the NS}
In stellar evolution code, we also consider the spin evolution of
pulsars as follows. With the spin-up of the NS, the accreting
material would interact with the magnetosphere of the NS. We
simply define the magnetosphere radius as the position that the
ram pressure of the infalling material is balanced by the magnetic
pressure of the NS \citep{lamb73}. Under assumption of spherical
accretion \citep{ghos79a,ghos79b}, the magnetosphere radius is
\begin{equation}
r_{\rm m}=1.6\times10^{8}\left(\frac{B_{\rm s}}{10^{12}\rm
G}\right)^{4/7}\left(\frac{|{\dot M}_{\rm d}|}{10^{18}{\rm g\,
s^{-1}}}\right)^{-2/7}{\rm cm},
\end{equation}
where $B_{\rm s}$ is the surface magnetic field of the NS. Some
observations and analysis argued that the mass accretion of the NS
can lead to its magnetic field decay \citep[see][]{wije97}. Here
we adopt an empirical model given by \cite{shi89}, i. e.
\begin{equation}
B_{\rm s}=\frac{B_{\rm i}}{1+\triangle M_{\rm acc}/m_{\rm B}}~,
\end{equation}
where $B_{\rm i}$ is the initial magnetic field of the NS,
$\triangle M_{\rm acc}$ is the accreted mass of the NS, and
$m_{\rm B}$ is $\sim10^{-4}M_\odot$.

When the NS rotation is too fast, the gravitational force of the
accreting material at $r_{\rm m}$ is less than its centrifugal
force. The centrifugal barrier would eject the accreting material
, and exerting a propeller spin-down torque on the NS
\citep{illa75}. Namely, if the magnetosphere radius is greater
than the co-rotation radius
\begin{equation}
r_{\rm c}=1.5\times10^8 \left(\frac{M_{\rm
NS}}{M_\odot}\right)^{1/3}P_{\rm s}^{2/3}~{\rm cm},
\end{equation}
where $P_{\rm s}$ is the spin-period of the NS in units of second,
the propeller effect occurs. The spin angular momentum loss rate
via the propeller effect can be written as
\begin{equation}
\dot{J}_{\rm p}=2\dot{M}r_{\rm m}^{2}[\Omega_{\rm K}(r_{\rm
m})-\Omega],
\end{equation}
where $\Omega_{\rm K}(r_{\rm m})$ is the Keplerian angular
velocity at $r_{\rm m}$. When $r_{\rm m}<r_{\rm co}$, the
accreting material is bound in the magnetic field lines to
co-rotate with the NS, and is accreted onto its surface. Assuming
rigid body rotation and the momentum of inertia $I=10^{45}\rm
g\,cm^{2}$, the spin-up torque of the accreting material exerting
on the NS is given by
\begin{equation}
\dot{J}_{\rm ac}=\dot{M}_{\rm ac}\sqrt{GM_{\rm NS}R},
\end{equation}
where $G$ is the gravitational constant, $R$ is the radius of the
NS.


In addition,  if $r_{\rm m}$ is greater than the light cylinder
radius
\begin{equation} r_{\rm {lc}}=\frac{
c}{\Omega}=\frac{ cP_{\rm s}}{2\pi},
\end{equation}
the NS appears as a radio pulsar. As a result of magnetic dipole
radiation, the spin angular momentum loss rate is
\begin{equation}
\dot{J}_{\rm m}=-\frac{2B^{2}_{\rm s}R^6\Omega^3}{3c^3}.
\end{equation}

\section{Results}
Based on the stellar evolution code and input physics described in
Section 2, we calculated the evolution of large numbers of LMXBs.
We take the initial mass of the donor star to be in the range $1.0
- 2.0~M_\odot$ (with a solar chemical composition Y = 0.28, Z =
0.02) \footnote{Based on the detailed numerical calculations for
the non-conservative evolution of close binaries, \citet{tau99}
concluded that binaries containing a donor star with mass of $\ga
2.0~M_\odot$ would experience a common envelope evolution, and
evolve into BMSPs with a short orbital period ($< 10~\rm days$)
and a heavy CO or ONeMg white dwarf.}, the initial mass of the NS
to be $1.4~M_\odot$. In addition, we assume that the NS is a dead
pulsar, which already evolved to cross the so-called death line,
and cannot radiate radio pulses. The spin-period of a dead pulsar
should satisfy $P_{\rm s}^2>B/1.7\times 10^{11}$ \citep{bhat92},
and the upper-limit of the spin-period for normal radio pulsars is
11 s \citep{man04}. \citet{wang11} proposed that the minimum spin
period of the NS is insensitive to its initial spin-period and
magnetic field. Therefore, we adopt an initial spin-period of
$P_{\rm s,i}=10~$ s and initial magnetic field of $B=10^{12}\rm G$
for the accreting NS. If the donor star evolves into a He white
dwarf and the Roche lobe overflow ends, we stop the calculation.

\begin{figure*}
 \includegraphics[width=\textwidth]{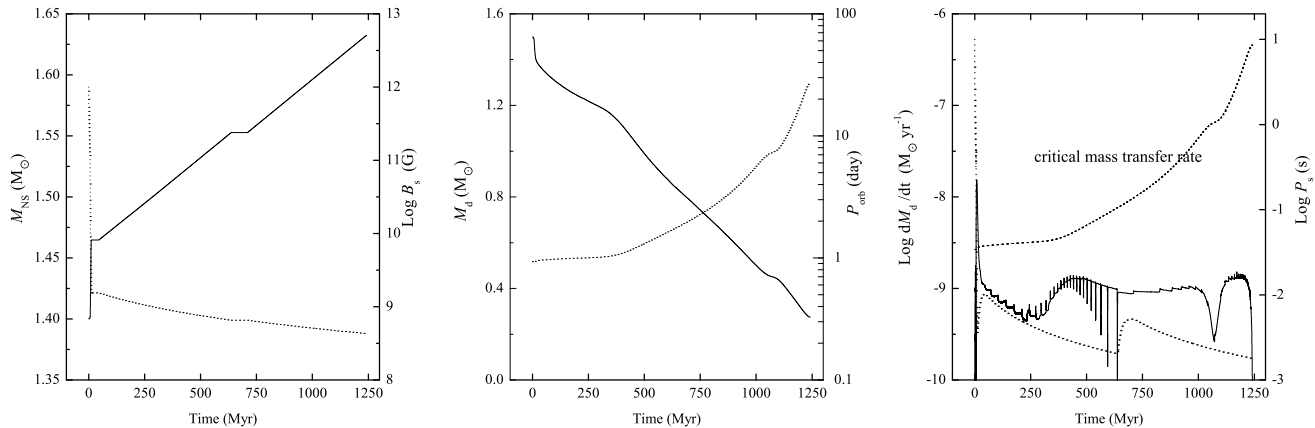}
  \caption{Evolutionary track of an LMXB with $M_{\rm d,i}$ = 1.5 $M_\odot$, and $P_{\rm orb,i}$ = 1.5 day, which can evolve into a BMSP. The solid and dotted curves represent the evolution
 the NS mass and the magnetic field in the left panel, the donor star mass
 and the orbital period in the middle panel, and the mass transfer rate and
 the spin period of the NS in the right panel, respectively.}
 \label{fig:fits}
\end{figure*}

In Figure 1, we show an evolutionary example of an LMXB with an
initial donor star of $M_{\rm d,i}$ = 1.5 $M_\odot$, and an
initial orbital period of $P_{\rm orb,i}$ = 1.5 day. Due to the
loss of orbital angular momentum by magnetic braking, the orbital
period decreases from 1.5 d to $\sim1.0$ d before the mass
exchange. When the age is $2.15\times10^9$ years, the donor starts
to overflow its Roche lobe, and mass transfer onto the NS
commences. Because the material transfers from the more massive
donor star to the less massive NS, the mass transfer firstly
occurs on a thermal timescale at a high rate of $\sim
10^{-8}~M_\odot\,\rm yr^{-1}$. At the same time, the surface
magnetic field of the NS sharp decay to $10^{9}$ G. In the initial
mass transfer phase, the LMXB appear as a short-lived persistent
X-ray source. With the decrease of the donor star mass, the mass
transfer subsequently occurs on a nuclear timescale at a lower
rate of $\sim 10^{-10}-10^{-9}~M_\odot\,\rm yr^{-1}$. This rate is
always less than the critical mass transfer rate \footnote{There
exist a lot of spikes and dips in the mass transfer rate curve,
which are the results that the donor star exhausted its core
hydrogen and deviated thermal equilibrium \citep{li04}.}.
Therefore, the LMXB should be a transient X-ray source about 95\%
of all its life. After $1.24\times10^9$ years mass transfer, the
NS grows to 1.63 $M_\odot$. By gaining the material and angular
momentum from the donor star, the spin-period $P_{\rm s}$ of the
NS continuously decrease to 1.8 ms. When the hydrogen-rich
envelope of the donor star is exhausted, the degenerate He-rich
core remains behind. The endpoint of the evolution is a BMBP
consisting a recycled pulsar and a He WD with a mass of
$0.27~M_{\odot}$, and with an orbital period of $27.3~\rm d$.

\begin{figure}
 \includegraphics[width=0.5\textwidth]{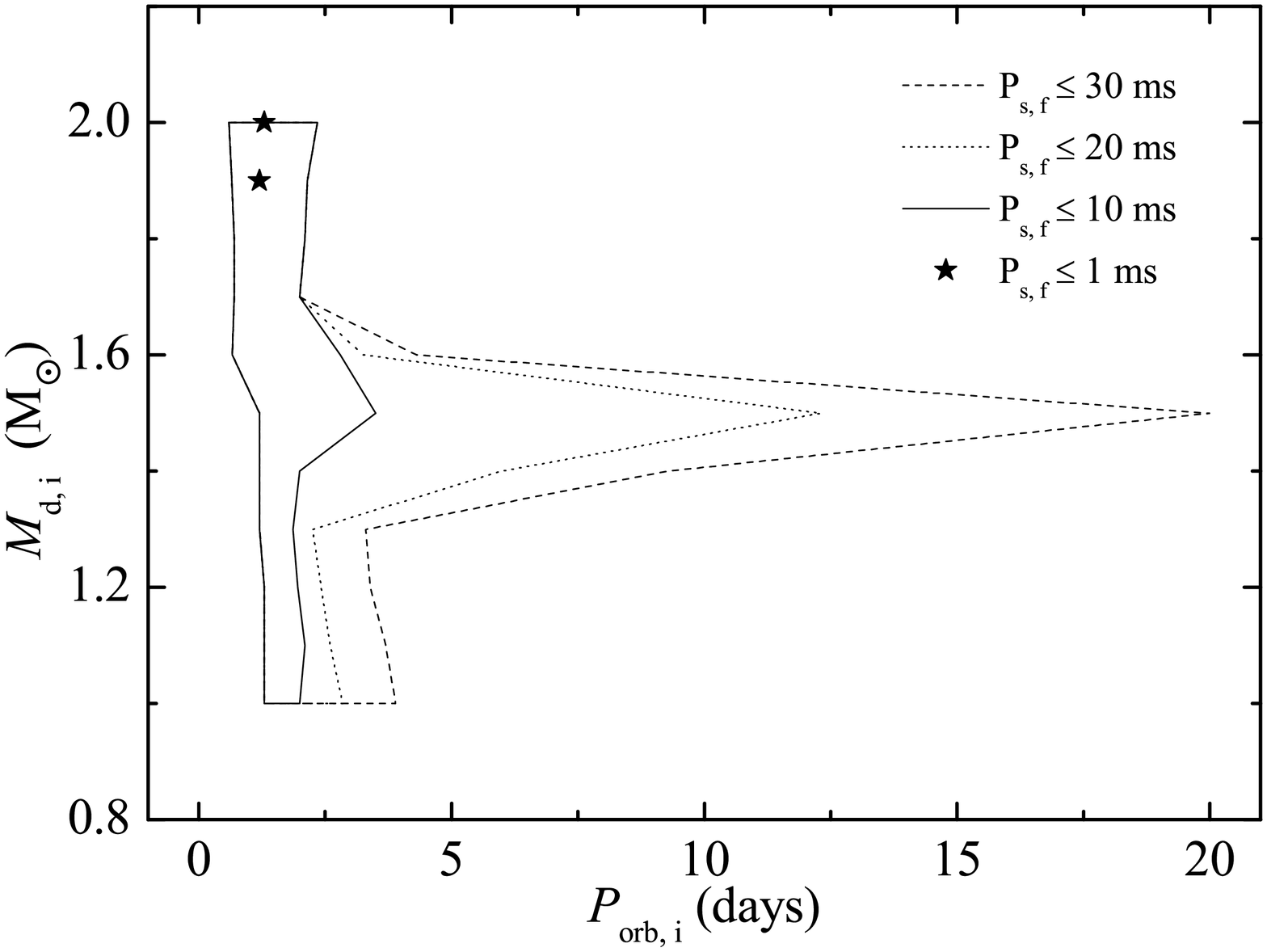}
  \caption{Distribution of the initial orbital periods $P_{{\rm orb,i}}$ and the initial donor star masses $M_{{\rm
  d,i}}$ of LMXBs that can evolve into BMSPs via the recycling evolutionary channel.}
 \label{fig:fits}
\end{figure}

The main aim of this work is to explore the progenitor properties
of BMSPs formed by the recycling evolutionary channel, therefore
we have calculated the evolution of large numbers of LMXBs with
different initial orbital periods and donor star masses. In Figure
2 we present the progenitor distribution of BMSPs in $M_{\rm
d,i}-P_{\rm orb,i}$ diagram. The regions enclosed by the solid,
dashed, and dotted curves represent the distribution areas of
LMXBs that can result in a BMSP with a spin period of 10 ms, 20
ms, and 30 ms, respectively. Our results show that all NSs in
LMXBs have a chance to be spun up to millisecond period, and the
final fate strong depend on the separation of the binary. When the
initial mass of the donor star is located in the range of 1.3
$M_\odot - 1.6~M_\odot$, the initial orbital period have a wider
distribution from 1.0 day to 20 days. When the initial mass of the
companion is between 1.0 $M_\odot$ and 1.4 $M_\odot$, the system
cannot produce a BMSP unless the initial orbital period is less
than 2.0 days. Beyond these areas, BMSPs cannot be formed due to
either a low spin-up efficiency or unstable mass transfer. For
donor stars with a mass of 1.4 -1.6 $M_\odot$, a lower mass
accumulation and spin-up efficiency of the NS result in an upper
limit of initial orbital period. However, for massive donor stars
with a mass of 1.7 -2.0 $M_\odot$, the upper limit on the orbital
period originates from the dynamical instability of mass transfer
\citep{will02}. In particular, in our calculated grids there exist
two binaries which can produce a sub-millisecond pulsar with a
spin period of 0.9 ms. Both LMXBs have a donor star with an
initial mass of 1.9 - 2.0 $M_\odot$, and an orbital period of
$1.2-1.3$ days.

\begin{figure}
 \includegraphics[width=0.5\textwidth]{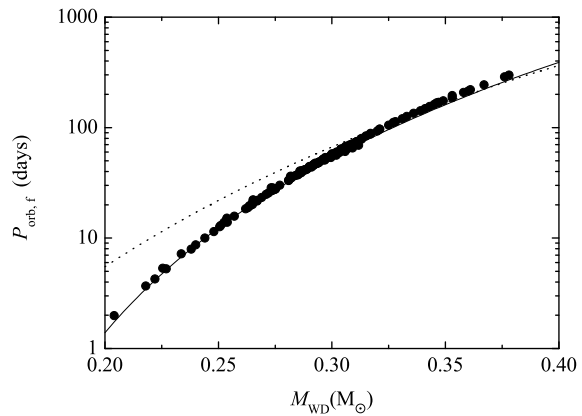}
  \caption{ Predicted relation between the orbital period $P_{\rm
  orb}$ and the white dwarf mass $M_{\rm WD}$ for low-mass binary pulsars.
 The filled circles, the solid curve and the dotted curve denote our calculated results, the relation obtained by \citet{tau99}, and \citet{rap95}, respectively.}
 \label{fig:fits}
\end{figure}

The evolutionary endpoint of most LMXBs is BMSPs consisting of a
millisecond pulsar and a low-mass He white dwarf, which is called
low-mass binary pulsar (LMBP) \citep{stai04,taur06}. Stellar
evolution theory predicts a tight relation between the core mass
of giants and their radius \citep{joss87}. During the evolution of
LMXBs, the giant should overflow its Roche lobe, and its radius
relates to the orbital separation. When the giant envelope is
exhausted, its core evolve into a white dwarf. Therefore, the
final orbital period of LMBPs should be correlated with the mass
of the white dwarf companion \citep{savo87,rap95}. Previous works
presented a simple relation between the orbital period $P_{\rm
orb}$ and the white dwarf mass $M_{\rm WD}$ for low-mass binary
pulsars (see also \citet{rap95} and \citet{tau99}). In Figure 3,
we show our obtained low-mass binary pulsars by filled circles in
$P_{\rm orb} - M_{\rm WD}$ diagram. It is clear that our
calculated results are consistent with the relation obtained by
\citet{tau99}. To compare with observations, we summarize the
observed parameters for 17 low-mass binary pulsars in Table 1. In
Figure 4, we compare the calculated results with the observed data
in the $ P_{\rm orb} - P_{\rm s}$ plane. It seems that our
evolutionary model can account for the formation of part BMSPs.
However, it is difficult for our evolutionary scenario to produce
BMSPs with a short spin-period (3-8 ms) and a long orbital period
($\ga 70-80$ day).

\begin{figure}
 \includegraphics[width=0.5\textwidth]{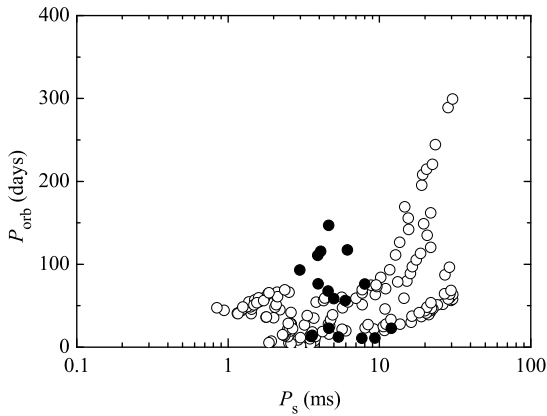}
  \caption{ Distribution of BMSPs in the spin-period $P_{\rm s}$ of BMSPs  vs. the final orbital periods $P_{\rm orb}$ diagram.
  The open circles, solid circles denote our calculated results and the observed data, respectively.}
 \label{fig:fits}
\end{figure}

In Figure 5, we show the distribution of the final accreted mass
and the final spin-period of NSs. One can see that, if NSs
accretes a mass of $\ga 0.1~M_{\odot}$, they can be spun up to
$\la 10~\rm ms$. In our calculated results, there exist 3 NSs that
can accrete mass of $\ga 0.6~M_{\odot}$. Recent Shapiro delay
measurements of PSR J1614-2230 suggested that it is a massive MSPs
($\sim 2~M_{\odot}$), and with a CO white dwarf of $\sim
0.5~M_{\odot}$ \citep{demo10} \footnote{Recently, \citet{lin11}
proposed that this MSP may originate from an IMXB with massive NS
of $1.6~M_{\odot}$. Another work performed by \citet{tau11} also
support the viewpoint that the NS in PSR J1614-2230 was born
massive.}. We expect the discovery of LMBP with a massive NS like
PSR J1614-2230 to test our evolutionary results.

\begin{table}
\begin{center}
\centering \caption{ Observed parameters for 17 low-mass binary
puslars.\label{tbl-1}}
\begin{tabular}{ccccc}
\hline\hline
Pulsars & $P_{\rm s}(\rm ms)$ &  $ P_{\rm orb}({\rm days})$ & $M_{\rm c}({\rm M}_\odot)$ & ${\rm References}$\\
\hline
J1455$-$3330 &  7.987 &  76.17 & 0.3 & 1\\
J1600$-$3053 &  3.598 &  14.35 & 0.2 & 2\\
J1618$-$3921 & 11.987 &  22.80 & 0.2 & 3\\
J1643$-$1224 & 4.622 &  147.02 & 0.1 & 1\\
J1709+2313 & 4.631 &  22.70 & 0.3  & 4\\
J1713+0747 & 4.570 &  67.83 & 0.3  & 5\\
J1751$-$2857 & 3.915 &  110.75 & 0.2 & 6\\
J1804$-$2717 & 9.343 &  11.13 & 0.2 & 7\\
J1853+1303 & 4.092 &  115.65 & 0.3  & 6\\
B1855+09   & 5.362 &  12.33 & 0.2  & 8\\
J1910+1256 & 4.984 &  58.47 & 0.2 & 6\\
J1918$-$0642 & 7.646 &  10.91 & 0.1 & 3\\
J1933$-$6211 & 3.543 &  12.82 & 0.4  & 2\\
B1953+29   & 6.133 &  117.35 & 0.2  & 9\\
J2019+2425 & 3.935 &  76.51 & 0.3 & 10\\
J2033+1734 & 5.949 &  56.31 & 0.2 & 11\\
J2229+2643 & 2.978 &  93.02 & 0.1  & 12\\
\hline\hline
\end{tabular}
   \begin{tablenotes}
     \item References: (1)\cite{lori95}; (2)\cite{jac07}; (3)\cite{edw01}; (4)\cite{lew04};
     (5)\cite{fos93}; (6)\cite{stai05}; (7)\cite{lori96}; (8)\cite{sege86}; (9)\cite{bori83}; (10)\cite{nic95}; (11)\cite{ray96}; (12)\cite{wol00}.
    \end{tablenotes}
\end{center}
\end{table}

\section{Discussion and summary}
Using a stellar evolution code, in this work we have investigated
the formation of BMSPs formed by the canonical recycling
evolutionary channel. In calculation, we take into account the
influence of thermal and viscous instability of an accretion disk
and propeller effect on the mass transfer process and the spin
evolution of NSs. Orbital angular momentum loss mechanism by
magnetic braking, which originate from the coupling between the
stellar winds and the magnetic field of the donor star, is also
considered. Our main results are summarized as follows.

\begin{figure}
 \includegraphics[width=0.5\textwidth]{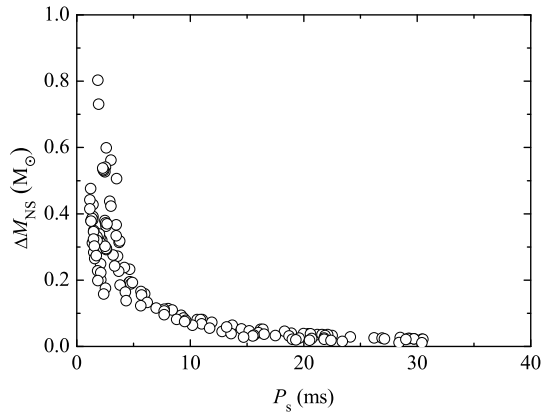}
  \caption{ Distribution of our simulated results in the spin-period $P_{\rm s}$ of BMSPs  vs. the accreted mass of NSs $\bigtriangleup M_{\rm NS}$ diagram.
  }
 \label{fig:fits}
\end{figure}

1. We obtain the initial donor star mass and orbital period
distribution of the progenitors of BMSPs formed by the recycled
channel (see Figure 2). Our results show that all LMXBs with a
donor star of 1.0 - 2.0 $M_\odot$ have a chance to evolve into a
BMSP with a spin-period of $P_{\rm s}\la \rm 10~ms$.

2. The final fate of LMXBs strongly depend on the initial donor
star mass and the separation of the binary. When the donor star
mass is in the range 1.0 - 1.4 $M_\odot$, only LMXBs with a short
orbital period ($P_{\rm orb}\la \rm 2.0~ day$) can evolve into
low-mass binary pulsars. However, for a higher mass donor star,
the upper limitation of the orbital period that can result in
birth of binary millisecond pulsars is 2.0 - 3.6 days.

3. Our calculated results show that, if the NS accretes a mass of
$\ga 0.1~M_\odot$, it can be spun up to millisecond period. In
addition, it is possible that few MSPs gain a mass of $\ga
0.5~M_\odot$.

4. It is difficult for our evolutionary scenario to produce a
sub-millisecond pulsars. This result is consistent with the
conclusion obtained by \citet{ferr07}. However, AIC evolutionary
channel may produce a sub-millisecond pulsar (or quark star)
\citep{du09}.

Obviously, our evolutionary results depend on the parameterized
input physics, especially the magnetic braking model, the duty
cycle, and the magnetic field decay model, which have not been
fully understood. Firstly, the loss of orbital angular momentum
plays a vital role in the evolution of LMXBs, hence magnetic
braking model can influence the final orbital period of BMSPs.
Secondly, the duty cycle can influence the outburst timescale and
the mass growth of the NS. Therefore, a large duty cycle can
result in the birth of MSPs with a short spin-period. In addition,
the duty cycle may relate to system parameters (Lasota 2001), and
may also evolve with the orbital period and mass transfer rate.
Thirdly, in our input physics the magnetosphere radius relates to
the field decay model, while their relation is not sensitive. Some
uncertainties mentioned above may be responsible for the
discrepancy between our simulated results and observational data
in Figure 4. A large duty cycle and a weak magnetic braking model
may produce BMSPs with a short spin period (3-8 ms) and a moderate
long orbital period ($\ga 70-80\rm d$). Certainly, if our
evolutionary model is correct, there may be other evolutionary
channel to BMSPs such as AIC process of massive white dwarfs.

\section*{Acknowledgments}
We would like to thank the anonymous reviewer for constructive
comments. This work was partly supported by the National Science
Foundation of China (No.10873011), Program for Science \&
Technology Innovation Talents in Universities of Henan Province,
and Innovation Scientists and Technicians Troop Construction
Projects of Henan Province, China.

\bsp

\label{lastpage}


\begin{thebibliography}{99}
\bibitem[\protect\citeauthoryear{Alexander \& Ferguson }{1994}]{alex94} Alexander D. R.,  Ferguson J. W. 1994, ApJ, 437, 879
\bibitem[\protect\citeauthoryear{Alpar et al.}{1982}]{alpa82} Alpar M. A., Cheng A. F., Ruderman M. A.,  Shaham J., 1982, Nat, 300, 728
\bibitem[\protect\citeauthoryear{Andronov et al.} {2003}]{andr03} Andronov N., Pinsonneault M.,  Sills A., 2003, ApJ, 582, 358
\bibitem[\protect\citeauthoryear{Archibald et al.} {2009}]{arch09} Archibald A. M., Stairs I. H., Ransom S. M., et al., 2009, Sci, 324, 1411
\bibitem[\protect\citeauthoryear{Bhattacharya \& van den Heuvel}{1991}]{bhat91} Bhattacharya D., van den Heuvel E. P. J., 1991, Phys. Rep., 203, 1
\bibitem[\protect\citeauthoryear{Bhattacharya et al.}{1992}]{bhat92} Bhattacharya D., Wijers R. A.M. J., Hartman J.W., Verbunt F., 1992, A\&A, 254, 198
\bibitem[\protect\citeauthoryear{Boriakoff et al.}{1983}]{bori83} Boriakoff V., Buccheri R., Fauci F., 1983, Nat, 304, 417
\bibitem[\protect\citeauthoryear{Chen \& Panei}{2011}]{chen11a} Chen W. -C., Panei J. A., 2011, A\&A, 527, A128
\bibitem[\protect\citeauthoryear{Chen et al.}{2011a}]{chen11c} Chen W. -C., Liu X. -W., Xu R. -X., Li X. -D., 2011a, MNRAS, 410, 1441
\bibitem[\protect\citeauthoryear{Chen et al.}{2011b}]{chen11b} Chen W. -C., Li X. -D., Xu R. -X., 2011b, A\&A, 530, A104
\bibitem[\protect\citeauthoryear{Dewi et al.}{2002}]{dewi02} Dewi J. D. M., Pols O. R., Savonije G. J., van den Heuvel E. P. J., 2002, MNRAS, 331, 1027
\bibitem[\protect\citeauthoryear{Demorest et al.}{2010}]{demo10} Demorest P. B., Pennucci T., Ransom S. M., Roberts M. S. E.,
Hessels J. W. T., 2010, Nat, 467, 1081
\bibitem[\protect\citeauthoryear{Du et al.}{2009}]{du09}Du Y. J., Xu R. X., Qiao G. J.,  Han J. L., 2009, MNRAS, 399, 1587
\bibitem[\protect\citeauthoryear{Dubus et al. }{1999}]{dubu99} Dubus G., Lasota J. -P., Hameury J. -M., Charles, P. 1999, MNRAS, 303, 139
\bibitem[\protect\citeauthoryear{Edwards \& Bailes }{2001}]{edw01} Edwards R.,  Bailes M., 2001, ApJ, 553, 801
\bibitem[\protect\citeauthoryear{Eggleton }{1971}]{egg71} Eggleton P. P., 1971, MNRAS, 151, 351
\bibitem[\protect\citeauthoryear{Eggleton }{1972}]{egg72} Eggleton P. P., 1972, MNRAS, 156, 361
\bibitem[\protect\citeauthoryear{Eggleton }{1973}]{egg73} Eggleton P. P., 1973, MNRAS, 163, 279
\bibitem[\protect\citeauthoryear{Ferrario \& Wickramasinghe}{2007}]{ferr07}Ferrario L., Wickramasinghe D., 2007, MNRAS, 375, 1009
\bibitem[\protect\citeauthoryear{Francischelli et al.}{2002}]{fra02} Francischelli G. J., Wijers R. A. M. J., Brown G. E., 2002, ApJ, 565, 471
\bibitem[\protect\citeauthoryear{Foster et al.}{1993}]{fos93}Foster R. S., Wolszczan A., Camilo F., 1993, ApJ, 410, L91
\bibitem[\protect\citeauthoryear{Ghosh \& Lamb}{1979a}]{ghos79a} Ghosh P., Lamb F. K., 1979a, ApJ, 232, 259
\bibitem[\protect\citeauthoryear{Ghosh \& Lamb}{1979b}]{ghos79b} Ghosh P., Lamb F. K., 1979b, ApJ, 234, 296
\bibitem[\protect\citeauthoryear{Han et al.}{1994}]{han94} Han Z., Podsiadlowski P., Eggleton P. P., 1994, MNRAS, 270, 121
\bibitem[\protect\citeauthoryear{Hurley et al.}{2010}]{hurl10} Hurley J. R., Tout C. A., Wickramasinghe D. T., Ferrario L., Kiel P. D., 2010, MNRAS, 402, 1437
\bibitem[\protect\citeauthoryear{Illarinov \& Sunyaev}{1975}]{illa75} Illarinov A. F., Sunyaev R. A., 1975, A\&A, 39, 185
\bibitem[\protect\citeauthoryear{Jacoby et al.}{2007}]{jac07} Jacoby B. A., Bailes M., Ord S. M., Knight H. S., Hotan A. W., 2007, ApJ, 656, 408
\bibitem[\protect\citeauthoryear{Jordan et al.}{2007}]{jord07} Jordan S., Aznar Cuadrado R., Napiwotzki R., Schmid H. M.,
Solanski S. K., 2007, A\&A, 462, 1097
\bibitem[\protect\citeauthoryear{Joss et al.}{1987}]{joss87} Joss P. C., Rappaport S., Lewis W., 1987, ApJ, 319, 180
\bibitem[\protect\citeauthoryear{Kim \& Demarque} {1996}]{kim96} Kim Y.-C., Demarque P., 1996, ApJ, 457, 340
\bibitem[\protect\citeauthoryear{King et al.}{1997}]{kin97} King A. R., Frank J., Kolb U., Titter H., 1997, ApJ, 484, 844
\bibitem[\protect\citeauthoryear{King et al.}{2003}]{kin03} King A. R., Rolfe D. J., Kolb U., Sshenker K., 2003, MNRAS, 341, L35
\bibitem[\protect\citeauthoryear{Konar \& Bhattacharya}{1997}]{kona97} Konar S., Bhattacharya D., 1997, MNRAS, 284, 311
\bibitem[\protect\citeauthoryear{Lasota}{2001}]{las01} Lasota J. -P., 2001, NewAR, 45, 449
\bibitem[\protect\citeauthoryear{Lamb et al.}{1973}]{lamb73} Lamb F. K., Pethick C. J., Pines D., 1973, ApJ, 184, 271
\bibitem[\protect\citeauthoryear{Lewandowski et al.}{2004}]{lew04}Lewandowski W., Wolszczan A., Feiler G., Konacki M; Soltysi$\rm {\acute{n}}$ski T., 2004, ApJ, 600, 905
\bibitem[\protect\citeauthoryear{Li}{2004}]{li04} Li X. -D, 2004, ApJ, 616, L119
\bibitem[\protect\citeauthoryear{Lin et al.}{2011}]{lin11} Lin J., Rappaport S., Podsiadlowski Ph., Nelson L., Paxton B., Todorov P., 2011, ApJ, 732, 70
\bibitem[\protect\citeauthoryear{Lorimer}{1995}]{lori95} Lorimer D. R., 1995, MNRAS, 274, 300
\bibitem[\protect\citeauthoryear{Lorimer et al.}{1996}]{lori96} Lorimer D. R., et al., 1996, MNRAS, 283, 1383
\bibitem[\protect\citeauthoryear{Lorimer}{2008}]{lori08} Lorimer D. R., 2008, Living Reviews in Relativity, 11, 8
\bibitem[\protect\citeauthoryear{Manchester}{2004}]{man04} Manchester R. N., 2004, Sci, 304, 542
\bibitem[\protect\citeauthoryear{Manchester et al.}{2005}]{man05} Manchester R. N., Hobbs G. B., Teoh A., Hobbs M. 2005, AJ, 129,
1993
\bibitem[\protect\citeauthoryear{Michel}{1987}]{mich87} Michel F. C., 1987, Nat, 329, 310
\bibitem[\protect\citeauthoryear{Nice \& Taylor}{1995}]{nic95} Nice D. J., Taylor J. H., 1995, ApJ, 441, 429
\bibitem[\protect\citeauthoryear{Nomoto }{1987}]{nomo87} Nomoto K., 1987, ApJ, 322, 206
\bibitem[\protect\citeauthoryear{Nomoto \& Kondo}{1991}]{nomo91} Nomoto K., Kondo, Y., 1991, ApJ, 367, L19
\bibitem[\protect\citeauthoryear{Podsiadlowski et al.}{2004}]{pods04} Podsiadlowski P., Langer N., Poelarends A. J. T., Rappaport S., Heger A., Pfahl E., 2004, ApJ, 612, 1044
\bibitem[\protect\citeauthoryear{Patterson }{1984}]{patt84} Patterson J., 1984, ApJS, 54, 443
\bibitem[\protect\citeauthoryear{Poelarends et al.} {2008}]{poel08} Poelarends A. J. T., Herwig F., Langer N., Heger, A., 2008, ApJ, 675, 614
\bibitem[\protect\citeauthoryear{Pols et al. }{1995}]{pol95} Pols O. R., Tout C. A., Eggleton P. P., Han Z., 1995, MNRAS, 274, 964
\bibitem[\protect\citeauthoryear{Pols et al. }{1998}]{pol98} Pols O. R., Schroder K. P., Hurley J. R., Tout C. A., 1998, MNRAS, 298, 525
\bibitem[\protect\citeauthoryear{Queloz et al. }{1998}]{quel98} Queloz D., Allain S., Mermilliod J. C., Bouvier J., Mayor M., 1998, A\&A, 335, 183
\bibitem[\protect\citeauthoryear{Ray et al.}{1996}]{ray96} Ray P. S., Thorsett S. E., Jenet F. A., van Kerkwijk M. H., Kulkarni S. R., Prince T. A., Sandhu J. S., Nice, D. J., 1996, ApJ, 470, 1103
\bibitem[\protect\citeauthoryear{Rappaport et al.} {1983}]{rapp83} Rappaport S., Verbunt F., Joss P. C., 1983, ApJ, 275, 713
\bibitem[\protect\citeauthoryear{Rappaport et al.}{1995}]{rap95} Rappaport S., Podsiadlowski Ph., Joss P. C., Stefano R. D., Han Z., 1995, MNRAS, 273, 731
\bibitem[\protect\citeauthoryear{Rogers \& Iglesias }{1992}]{roge92} Rogers F. J., Iglesias C. A., 1992, ApJS, 79, 507
\bibitem[\protect\citeauthoryear{Savonije }{1987}]{savo87} Savonije G. J., 1987, Nat, 325, 416
\bibitem[\protect\citeauthoryear{Segelstein et al.}{1986}]{sege86} Segelstein D. J., Rawley L. A., Stinebring D. R., Fruchter A. S., Taylor, J. H., 1986, Nat, 322, 714
\bibitem[\protect\citeauthoryear{Shibazaki et al.}{1989}]{shi89} Shibazaki N., Murakami T., Shaham J., Nomoto K., 1989, Nat, 342, 656
\bibitem[\protect\citeauthoryear{Siess} {2007}]{sies07} Siess, L. 2007, A\&A, 476, 893
\bibitem[\protect\citeauthoryear{Sills et al.}{2000}]{sill00} Sills A., Pinsonneault M. H., Terndrup D. M., 2000, ApJ, 534, 335
\bibitem[\protect\citeauthoryear{Stairs}{2004}]{stai04} Stairs I. H., 2004, Sci, 304, 547
\bibitem[\protect\citeauthoryear{Stairs}{2005}]{stai05} Stairs I. H., 2005, ApJ, 632, 1060
\bibitem[\protect\citeauthoryear{Tauris \& Savonije}{1999}]{tau99} Tauris T. M., Gerrit J. Savonije., 1999, A\&A, 350, 928
\bibitem[\protect\citeauthoryear{Tauris \& van den Heuvel}{2006}]{taur06} Tauris T. M., van den Heuvel E. P. J., 2006, in Compact stellar
X-ray sources. ed. by W. Lewin \& M. van der Klis (Cambridge:
Cambridge Univ. Press), 623
\bibitem[\protect\citeauthoryear{Tauris, Langer \& Kramer}{2011}]{tau11} Tauris T. M., Langer N., Kramer M., 2011, MNRAS, in press [arXiv:1103.4996]
\bibitem[van Paradijs (1996)]{para96} van Paradijs, J. 1996, ApJ, 464, L139
\bibitem[Verbun \& Zwaan (1981)]{verb81} Verbunt F., Zwaan C., 1981, A\&A, 100, L7
\bibitem[\protect\citeauthoryear{Wang et al. }{2011}]{wang11} Wang J., Zhang C. M., Zhao Y. H., Kojima Y., Yin H. X., Song L.
M., 2011,  A\&A, 526, A88
\bibitem[\protect\citeauthoryear{Wickramasinghe et al.}{2009}]{wic09} Wickramasinghe D. T., Hurley J. R., Ferrario L., Tout C. A., Kiel P. D., 2009, JPhCS, 172, 2037
\bibitem[\protect\citeauthoryear{Wijers}{1997}]{wije97} Wijers R. A. M. J., 1997, MNRAS, 287, 607
\bibitem[\protect\citeauthoryear{Wijnands \& van der Klis}{1998}]{wijn98} Wijnands R., van der Klis M., 1998, Nat, 394, 344
\bibitem[\protect\citeauthoryear{Willems \& Kolb}{2002}]{will02} Willems B., Kolb, U., 2002, MNRAS, 337, 1004
\bibitem[\protect\citeauthoryear{Wolszczan et al.}{2000}]{wol00} Wolszczan A., Doroshenko O. V., Konacki M., Kramer M., Jessner A., Wielebinski R., Camilo F., Nice D. J., Taylor J. H., 2000, ApJ, 528, 907



\end{thebibliography}
\end{document}